\documentclass{iopart}
\usepackage{iopams}  
\usepackage{graphicx}
\usepackage{color}
\newcommand{\bm}[1]{\mbox{\boldmath$#1$}}
\newcommand{\be}{\begin{equation}}
\newcommand{\ee}{\end{equation}}
\newcommand{\bea}{\begin{eqnarray}}
\newcommand{\eea}{\end{eqnarray}}
\newcommand{\non}{\nonumber}
\newtheorem{theorem}{Theorem}[section]

\newtheorem{lemma}[theorem]{Lemma}

\begin{document}
\title[Exact quantum numbers of collapsed 2-string solutions]
{Exact quantum numbers of collapsed and non-collapsed 2-string solutions in the Heisenberg spin chain } 
\author{Tetsuo Deguchi and Pulak Ranjan Giri }

\address{Department of Physics, Natural Science Division, Faculty of Core Research, 
Ochanomizu University \\
2-1-1 Ohtsuka, Bunkyo-ku, Tokyo 112-8610, Japan}

\eads{\mailto{deguchi@phys.ocha.ac.jp}}

\date{\today}

\begin{abstract}
Every solution of the Bethe-ansatz equations (BAE) is characterized by a  set of quantum numbers, by which we can evaluate it numerically. However, no general rule  is known how to give quantum numbers for the physical solutions of  BAE. 
For the spin-1/2 XXX chain we rigorously derive all the quantum numbers for the complete set of  the Bethe-ansatz eigenvectors in the two down-spin sector  with any chain length $N$. Here we obtain them both for real and complex  solutions. 
Consequently,  we prove the completeness of the Bethe ansatz  and give an exact expression for the number of real solutions which correspond to collapsed bound-state solutions (i.e., 2-string solutions)  in the sector: $2[ (N-1)/2 - (N/\pi) \tan^{-1}(\sqrt{N-1})]$ in terms of Gauss' symbol.  Moreover,  we prove in the sector the scheme conjectured by Takahashi  for solving BAE systematically.  
We also suggest that by applying the present method  we can derive the quantum numbers for the spin-1/2  XXZ chain.   
\end{abstract}

\maketitle

\setcounter{equation}{0} 
\renewcommand{\theequation}{1.\arabic{equation}}
\section{Introduction}

Exact solutions of the one-dimensional Heisenberg model, i.e. the spin-1/2 XXX chain, 
were derived by Bethe in 1930s \cite{bethe} with a systematic method by assuming the form of wavefunctions.  We  call it the coordinate Bethe ansatz \cite{gaudin}. 
After Baxter solved the XYZ spin chain, i.e. the eight-vertex model,  
by introducing and solving the Yang-Baxter equation \cite{Baxter}, 
powerful algebraic methods such as the algebraic Bethe ansatz have been developed \cite{faddeev-takhtajan,Sklyanin,korepin,faddeev1,suth}, where the Yang-Baxter equation leads to algebraic relations such as the commutation relations among the operator elements of the monodromy matrix. 
However, it is still not trivial whether the eigenvectors constructed by the Bethe-ansatz  methods are complete \cite{bethe,koma}.  
In the Bethe ansatz each Bethe-ansatz eigenvector corresponds to a solution of the Bethe-ansatz equations (BAE).  A useful but conjectured scheme was formulated to evaluate 
all physical solutions of BAE \cite{bethe,taka,takab}. 
We call it the {\it string hypothesis}. 
In order to confirm it all the solutions are numerically evaluated
for the $N$-sited XXX spin chains with $N=10, 12$ and 14 \cite{hag,nepo1,giri3} 
(see also \cite{kirillov}).

According to the string hypothesis any solution to BAE is given by either a real number or a sequence of complex numbers, depending on the quantum numbers \cite{taka}.  We call such a sequence of complex numbers a {\it string}.   
However, it has been shown by Essler, Korepin and Schoutens that some solutions with two down-spins which should be complex in the string hypothesis are indeed real if the system is larger than some critical size \cite{essler}.  
We call  it  the collapse of  a potentially complex or bound-state solution to a real solution. 
The collapse  of such a supposed to be complex solution in the string hypothesis 
to a real solution  was observed  numerically also in other sectors \cite{hag}. 
Thus, in order to study the completeness of the Bethe ansatz,  it is fundamental to show whether the number of  such new real solutions that appear after 
some complex solutions collapse  corresponds to the number of  the missing complex solutions.

It is known that every solution of BAE is characterized by a  set of quantum numbers. 
If the quantum numbers of a complex solution are known, we can numerically evaluate its deviations from the ideal form of strings even though they are very small \cite{giri3}.  However, no general rule  is known how to give quantum numbers for the physical solutions of BAE.  It is thus quite important to determine the quantum number of any physical solution of BAE exactly. 

In the present paper we rigorously derive all the quantum numbers for a complete set of the Bethe-ansatz eigenvectors in the sector of two down-spins in the spin-1/2 XXX chain of any chain length $N$.  We first recall that any complex solution with two down-spins is expressed in terms of two real parameters \cite{vlad2}.  We solve BAE with respect to one of the two parameters and introduce functions which give quantum numbers as their special values. We show that they are monotonic and give the upper and lower bounds to them analytically.  We thus obtain the quantum numbers  both for real and complex solutions in the sector  without making any assumption.    
Consequently,  we prove the completeness of the Bethe-ansatz in the two down-spin sector  for any system size $N$ analytically via the Bethe-ansatz equations. 
Furthermore,  we derive an exact expression for the number of  missing complex solutions in the sector and that of new real solutions generated after the collapse of complex solutions occurs. We observe that they are equal to each other  
in  the sector for any chain length $N$:  
$2[ (N-1)/2 - (N/\pi) \tan^{-1}(\sqrt{N-1})]$ in terms of Gauss' symbol.   
Here we remark that the formula has also been obtained independently by Caux \cite{Caux}.  
Furthermore, we prove the scheme for solving all complex solutions conjectured by Takahashi rigorously in the two down-spin sector.   In particular, we derive the relation of the Takahashi quantum numbers to the Bethe quantum numbers, which we shall define, shortly in Introduction. Here we make an only one assumption in the paper that   
the deviations of complex solutions from the complete form (i.e. the string deviations) are very small when we define the Takahashi quantum number. 
However,  the relation between the Bethe and Takahashi quantum numbers is valid even when the deviations are not very small.   
Based on Ref. \cite{giri3} we suggest that the result could be related to the characterization of  solutions of BAE  with rigged configurations \cite{kirillov,kirillov1,kirillov3,kirillov4,lule}.   
Moreover,  we confirm some features of complex solutions analytically  such as singular string solutions \cite{vlad1,sid,bei,nepo,nepot,nepoh,kirillov2,giri1,giri2} and the large-$N$ behavior of complex solutions \cite{vlad,isler,ila,fuji}.  Finally, we suggest that we can derive the quantum numbers of  complex solutions with two down-spins for the spin-1/2 XXZ chain, for instance,  in the gapful antiferromagnetic regime, by applying the method in the present paper. There are earlier studies on the string solutions of the XXZ spin chain \cite{woy,bab,fab1,fab2}.

Let us introduce the Bethe-ansatz eqautions  for the XXX spin chain. 
In  the $M$ down-spin sector they are given  for  $M$ rapidities $\lambda_1, \lambda_2, \ldots,  \lambda_M$  in the multiplicative form as follows.  
\be 
\left( \frac {\lambda_{\alpha} + i/2}  {\lambda_{\alpha} - i/2} \right)^N  
= \prod_{\beta=1;  \beta \ne \alpha}^{M}  
\frac {\lambda_{\alpha}- \lambda_{\beta} + i} 
 {\lambda_{\alpha}- \lambda_{\beta} - i} \, ,  \quad 
\mbox{ for} \, \, \alpha= 1, 2 ,\ldots, M \, . 
\label{eq:BAEmul}
\ee
By taking the logarithm of the both hand sides of (\ref{eq:BAEmul}) 
 and expressing the logarithmic function of a complex argument 
in terms of the arctangent function as shown in eq. (\ref{eq:log-z}) of Appendix A 
we have 
\bea 
2 \tan^{-1} \left(  2 \lambda_{\alpha} \right) & = & {\frac {2 \pi} N}   J_{\alpha} 
+ {\frac 1 N} \sum_{\beta=1}^{M} 2 \tan^{-1} 
\left( \lambda_{\alpha} - \lambda_{\beta} \right), \non \\ 
& &  \quad  \mbox{ for} \, \, \alpha= 1, 2 ,\ldots, M .  
\label{eq:BAElog}
\eea 
Here we call $J_{\alpha}$ the Bethe quantum numbers \cite{hag}. 
They are given by integers or half-integers according to the condition 
\be 
J_{\alpha} = {\frac 1 2}(N-M+1)  \quad  ({\rm mod}  \, 1)  \, . 
\label{eq:parityJ}
\ee  
In the paper we take the branch of the arctangent function:  
$- \pi/2 < \tan^{-1} x < \pi/2$ for any $x \in {\bm R}$, and hence       
we do not assume any additional integral multiple of  $2 \pi$ in (\ref{eq:BAElog}).

In the string hypothesis we assume that  any solution to the Bethe-ansatz equations
(\ref{eq:BAEmul}) is given by a sequence of $n$ complex numbers which are different from  each other by integral  multiples of an imaginary number such as $i$, at least approximately: 
\be 
\lambda_{\alpha k}^n = x_{\alpha}^n +  {\frac i 2} (n +1  - 2 k) 
+ \Delta_{\alpha k}^n \quad \mbox{for} \, \, 
k= 1, 2, \ldots, n.    \label{eq:n-string}
\ee
We call the set of rapidities of the form (\ref{eq:n-string}) an $n$-string.   
We call   $x_{\alpha}^n$ the center of the $n$-string and $\Delta_{\alpha k}^n$ 
the deviations from the complete $n$-string. 
Here, suffix $\alpha$  being the index to distinguish  all the $M_n$  strings of 
the same length $\Delta_{\alpha k}^{n}$  are  the string deviations.  
We remark that a real solution is given by $1$-strings, which are given 
by putting $n=1$ in  (\ref{eq:n-string}). 
In the string hypothesis we assume that 
the absolute values of  the deviations are very small: $|\Delta_{\alpha k}^{n}| \ll 1$.  
In the limit that these  deviations vanish,  $\Delta_{\alpha k}^{n} \to 0$, 
the Bethe-ansatz equations  reduce to the convenient form
\begin{eqnarray} \label{betheta} 
\nonumber
\tan^{-1} \left( \frac{2\lambda^n_\alpha}{n} \right) &=& \pi \frac{I^n_\alpha}{N} + \frac{1}{N}\sum_{k=1}^{N_s}\sum_{\beta=1}^{M_k} 
\Theta_{nk}\left(\lambda_\alpha^n-\lambda_\beta^k\right)\,,~~~~ \\
\Theta_{jk}(\lambda) &=& (1-\delta_{nk}) \tan^{-1} \left( \frac{2\lambda}{|n-k|} \right) 
+ \tan^{-1} \left( \frac{2\lambda}{|n-k|+2} \right) + \cdots 
\non \\ 
& & + \cdots + 2\tan^{-1} \left( \frac{2\lambda}{n+k-2} \right)
+ \tan^{-1} \left( \frac{2\lambda}{n+k} \right) \,, 
\end{eqnarray}
where  $M$ down-spins are partitioned into  $M_k$ $k$-stings with the length of the largest string being  $N_s$  such that  $\sum_{k}^{N_s} kM_k=M$.  The reduced equations 
(\ref{betheta}) are called the Takahashi (or the Bethe-Takahashi) equations.   
 We now obtain the strictly non-repetitive quantum numbers   $I^j_\alpha$, known as the Takahashi quantum numbers,  from eq.   (\ref{betheta}), which have the following bounds
\begin{eqnarray}\label{takahashi} 
\mid I_{\alpha}^j\mid \leq \frac{1}{2} \left(N-1-\sum_{k=1}\left[2 \mbox{min}(j,k)-\delta_{j,k}\right]M_k\right)\,.
\end{eqnarray}

We recall that it has not been clearly shown previously how to derive all the possible Bethe quantum numbers for  the Bethe-ansatz equations (\ref{eq:BAElog}) with $M$ down-spins even for the $M=2$ case. Furthermore,  we observe in some examples that the same numbers appear in the Bethe quantum numbers for different rapidities, i.e. the Bethe quantum numbers are repetitive.  However, if we assume the string hypothesis then  
the Takahashi quantum numbers (or the Bethe-Takahashi quantum numbers \cite{hag}) are not repetitive,  and hence are indeed useful for not only classifying  but also evaluating solutions to the Bethe-ansatz equations numerically.

The contents of the paper consist of the following. 
In section 2 we derive rigorously the Bethe quantum numbers for 2-string solutions in the 
 two down-spin sector.  In section 3 we derive rigorously 
the Bethe quantum numbers for 1-string solutions in the 
 two down-spin sector.  Here we extend the parameters of  complex solutions to those of a pair of real solutions.  
In section 4 we show some features of 2-string solutions 
with two down-spins.  In section 5 we show analytically 
the relation between the Bethe quantum numbers and the Takahashi quantum numbers.
We show that the Takahashi quantum numbers are distinct, while the Bethe quantum numbers may be repetitive although they are distinct as pairs.  In section 6 we give some remarks on how far we have shown the completeness of the Bethe ansatz, and on suggestions for further studies.

%%%%%%%%%%%%%%%%%%%%%%%%%%%%%%%
%  Sec 2
%
\setcounter{equation}{0} 
\renewcommand{\theequation}{2.\arabic{equation}}
\section{Bethe quantum numbers of the complex solutions with two down-spins}

\subsection{Assumed form of a 2-string solution}

Le us recall that every physical solution  is self-conjugate under complex conjugation  in the XXX spin chain \cite{vlad2}. 
We therefore assume the following form of a 2-string solution   
to the Bethe-ansatz equations (\ref{eq:BAElog})  in the two down-spin sector ($M=2$) : 
\bea
\lambda_1 & = & x + {\frac i 2} (1 + 2 \delta) \, ,  \non \\ 
\lambda_2 & = & x - {\frac i 2} (1 + 2 \delta) \, . 
\eea
Here we assume that both the string center $x$ and the string deviation $\delta$ are real.   
Moreover, we assume that $-1/2 \le \delta$ due to the symmetry between 
$\lambda_1 $ and $\lambda_2$.   We shall consider two regimes of  $\delta$:  the positive regime with $\delta > 0 $ and the negative regime satisfying 
$-1/2  <  \delta < 0$.  
The Bethe-ansatz equations in the logarithmic form for a 2-string  
with $M=2$  are explicitly given by 
\bea 
2 \tan^{-1} \left(  2x + i(1+ 2 \delta) \right)  & = & {\frac {2 \pi} N}   J_{1} 
+ {\frac 1 N}  2 \tan^{-1} \left( i(1+ 2\delta) \right) \, ,  \label{eq:BAE1}
\\ 
2 \tan^{-1} \left( 2x - i(1+ 2 \delta) \right)  & = & {\frac {2 \pi} N}   J_{2} 
+ {\frac 1 N}  2 \tan^{-1} \left( - i (1+ 2 \delta) \right) \, .  \label{eq:BAE2}
\eea

Let us now solve the Bethe-ansatz equations (\ref{eq:BAE1}) and (\ref{eq:BAE2})  analytically without making any approximation.  
We introduce the step function $H(x)$ by 
\be 
H(x) = \left\{ 
\begin{array}{cc}  
1 & \mbox{for} \, \, x > 0 \, \\ 
0 & \mbox{ otherwise .}   
\end{array} \right. 
\ee
We also define sign functions  ${\rm sgn}(x_{\pm})$  by 
${\rm sgn}(x_{+}) = 1 - 2 H(-x)$ and ${\rm sgn}(x_{-}) =  2 H(x)  - 1 $, respectively.  
By applying  the expression of the arctangent function with a complex argument 
given in  eq. (\ref{eq:arctan})  we calculate  the left-hand side of (\ref{eq:BAE1}) as follows.  
\bea 
& & 2 \tan^{-1}(2x + i (1+ 2\delta))  =   
 \tan^{-1} \left( \frac x {1+ \delta}  \right) -  \tan^{-1} \left(  \frac x {\delta} \right)   
 \non \\ 
& & \qquad + \pi H(\delta) {\rm sgn}(x_{+})  + \frac 1 {2 i} \log \left( \frac {x^2 + \delta^2} {x^2 + (1+ \delta)^2}   \right) \, .  \label{eq:x-d}
\eea 
Here we have considered $H(-2 - 2\delta) = 0$  since $\delta > -1/2$. 
Moreover, in the paper we assume the branch of logarithmic function $\log z$ satisfying $|\Im \log z| \le \pi$, which is given by setting $n=0$ in (\ref{eq:log-z}), if we do not specify it otherwise.   

Let us define counting functions $Z_{\pm}(x, \delta)$, 
which are functions of the two variables, string center $x$ and string deviation $\delta$, by 
\bea 
2 \pi Z_{\pm }(x, \delta)  & = &  \pi H(\delta) \left({\rm sgn}(x_{\pm}) - (\pm 1) \frac 1 N \right) \non \\ 
& &  + \tan^{-1} \left( \frac x {1+ \delta}  \right)
 -  \tan^{-1} \left(  \frac x {\delta} \right) \, .    
\label{eq:Z}  
\eea
Through (\ref{eq:x-d}) we show that the first BAE (\ref{eq:BAE1}) holds  
if the following equations are satisfied:  
\bea 
& & {\frac {2 \pi } N}  J_1  =  2 \pi Z_{+}(x, \delta) \, ,   \label{BAE1Re} \\ 
&& \log \left( \frac {x^2 + \delta^2} {x^{2} + (1+\delta)^2} \right) = 
{\frac 1 N} \log \left( {\frac {\delta^2} {(1 + \delta)^2} } \right)   \, .  \label{BAEim}
\eea 
Applying formula (\ref{eq:x-d}) to the second BAE (\ref{eq:BAE2}), where we only replace 
$i$ with $-i$,  
 we show that   
it holds if eq. (\ref{BAEim}) and the following equation are satisfied: 
\be 
  {\frac {2 \pi } N}  J_2  =  2 \pi Z_{-}(x, \delta)  \, .    
\label{BAE2Re} 
\ee 
Thus, BAEs (\ref{eq:BAE1}) and (\ref{eq:BAE2}) are derived from  
the three equations (\ref{BAE1Re}),  (\ref{BAEim}) and (\ref{BAE2Re}), 
and if we take the branch of $\log z$ with $|{\rm Im} \log z| \le \pi$, 
they  are equivalent to the BAEs.

For the 2-string solutions the Bethe quantum  numbers $J_1$ and $J_2$ are not independent of each other.   By taking the difference of (\ref{BAE1Re}) and (\ref{BAE2Re}) for $x \ne 0$  we have the relation between them.  
\be 
J_2 = J_1 + H(\delta) \, . 
\ee

We now express the square of string center $x^2$ as a function of string deviation $\delta$. 
By taking the exponential of  (\ref{BAEim}) we have 
\be 
 \frac {x^2 + \delta^2} {x^{2} + (1+\delta)^2} = 
\left( {\frac {\delta^2} {(1 + \delta)^2} } \right)^{1/N} \, . \label{eq:x^2-del}
\ee
 We therefore obtain the following expression of  $x^2$ as a function of $\delta$:    
\bea 
x^2 & = & (1+ \delta)^2 
\frac { \left({\delta^2}/ {(1+\delta)^2} \right)^{1/N} -  {\delta^2}/{(1+\delta)^2} }
  {1 -  \left( {\delta^2}/{(1+\delta)^2} \right)^{1/N} } \, . 
\label{eq:x^2}
\eea

%%%%%%%%%%%%%%%%%%%%%%%%%%%%%
\subsection{Cases of the string deviation being real}

In order to parametrize string deviation $\delta$  
we introduce variable $\beta$ by  
\be 
\delta = - {\frac 1 2} + {\frac {\beta} 2} \qquad (\beta \ge 0)  . 
\label{eq:beta}
\ee
For positive $\delta$ ($\delta > 0$)  we have  $\beta > 1$, while 
for negative $\delta$ ($-1/2  <  \delta < 0$)  we have  $0 < \beta < 1$.  

We define an important variable $w$ by  
\be 
w= {\frac {\delta^2} {(1+\delta)^2}}  \, . 
\ee 
In terms of $\beta$ we have $w= (\beta -1)^2/(1+ \beta)^2$. 
For $\delta \ge 0 $, variable $w$ increases from 0 to 1 as $\beta$ increases from 1 to $\infty$.  For $-1/2  <  \delta \le 0$, variable $w$ increases from 0 to 1 
as $\beta$ decreases from 1 to 0. We shall show later 
that the 2-string solution collapses when 
$\beta$ approaches 0 ($\delta=-1/2)$.

It follows from (\ref{eq:x^2}) that ratios $x^2/{(1+ \delta)^2}$ and $x^2/{\delta}^2$ 
are expressed as functions of single variable $w$. 
\be  
\frac {x^2} {(1+ \delta)^2} =  \frac {w^{1/N} - w} {1-w^{1/N}} \, , \quad  
\frac {x^2} {\delta^2} = \frac {w^{1/N-1} - 1} {1-w^{1/N}}  \, . \label{eq:ratios}
\ee
We now consider four cases with respect to string deviation $\delta$ and 
string center $x$: (i) $\delta >$ 0 and $x> 0$;   (ii) $\delta >$ 0 and $x< 0$;  
(iii) $\delta <$ 0 and $x> 0$;    (iv) $\delta <$ 0 and $x < 0$.   
By taking the square roots of  $x^2/{(1+ \delta)^2}$ and $x^2/{\delta}^2$ 
ratios  $x/{(1+ \delta)}$ and $x/{\delta}$ are expressed  
in terms of $\epsilon_1= \pm 1$ and  $\epsilon_2= \pm 1$ as 
\be 
{\frac  x {1+ \delta}}(w) = \epsilon_1 \sqrt{ \frac {w^{1/N} - w} {1-w^{1/N}} } \, , \quad 
{\frac  x {\delta}}(w) = \epsilon_2 \sqrt{ \frac {w^{1/N-1} - 1} {1-w^{1/N}} } \, . 
\label{eq:xx}
\ee
In terms of signs $\epsilon_1$ and $\epsilon_2$ in  (\ref{eq:xx}) 
we specify cases (i), (ii), (iii) and (iv) with $(\epsilon_1, \epsilon_2) = (+, +)$ , $(-, -)$, $(+, -)$ and $(-, +)$, respectively.  
Consequently, we have  ${\delta}/(1+ \delta) = \epsilon_1 \epsilon_2 w^{1/2}$ 
and we can express $\delta$ as a function of $w$. 
\be 
\delta = {\frac 1 {1- \epsilon_1 \epsilon_2 w^{1/2}} } - 1 \, .  \label{eq:delta-w}
\ee
We define $\Theta(w)$  by 
\be 
\Theta(w) =  \tan^{-1}\left( \frac x {1+ \delta} (w) \right)
- \tan^{-1}\left( \frac x {\delta}(w) \right) \, . 
\ee 
Thus, by solving the constraint (\ref{BAEim}) in terms of $w$, 
we express the counting functions $Z_{\pm}(x, \delta)$ as 
$Z_{\pm}(x(w), \delta(w))$  as functions of $w$ for $0 \le w \le 1$. 
\be 
2 \pi Z_{\pm }(x(w), \delta(w))   =  \Theta(w) + 
\pi H(\delta(w) ) \left({\rm sgn}(x(w)+ 0_{\pm}) - (\pm 1) \frac 1 N \right) . 
\ee

Through direct calculation we can show   
\bea 
{\frac {d \Theta} {d w}} & = & \frac 
{\epsilon_1 (1-w) (w^{1/N-1/2} -1)  + \epsilon_2 N (1- w^{1/2}) (1- w^{1/N})}  
{2N (1-w) \sqrt{(1-w^{1/N})(w^{1/N-1} -1) } } .  \label{eq:dTheta} 
\eea
It is easy to show that 
\be 
(1-w) (w^{1/N-1/2} -1) + N (1- w^{1/2}) (1- w^{1/N}) > 0  \quad \mbox{ for} 
\, \,  0 < w < 1.  \label{eq:ineq1}
\ee
Furthermore, we can show the following lemma (the proof is given in Appendix B). 
\begin{lemma} 
For $0 < w < 1$ we have 
\be 
(1-w) (1 + w^{1/N-1/2}) - N (1+ w^{1/2}) (1- w^{1/N}) > 0 \, .  
\label{eq:ineq2}
\ee
\label{lem:2}
\end{lemma}

It follows from (\ref{eq:dTheta}), (\ref{eq:ineq1}) and  (\ref{eq:ineq2}) 
that we have either $d \Theta/dw > 0$ or 
$d \Theta/dw < 0$ in each of the four cases from (i) to (iv). 
If  $d \Theta/dw > 0$,   the lower bound (or the minimum) and the upper bound  of  function  $\Theta(w)$ are given by the limiting value of $\Theta(w=0)$ when we send $w$ to zero with $w > 0$   and that of  $\Theta(w = 1)$ when we send $w$ to 1, respectively.   If  $d \Theta/dw  < 0$,     the upper bound  and the lower bound  (or the minimum)  of  function  $\Theta(w)$ are given by 
the limiting value of $\Theta(w=0)$ when we send $w$ to zero with $w > 0$
and that of $\Theta(w=1)$ when we  send $w$ to 1, respectively.

By making use of the positivity (or negativity) of the derivative of 
the counting function $Z_{\pm}(x(w), \delta(w))$ 
with respect to $w$  we have the following results.  
\par \noindent 
Case (i):  $\delta > 0$ (i.e., $1 < \beta$) and $ x> 0$  
\be 
{\frac N 4} - {\frac 1 2} \le J_1 < {\frac N 2 } - {\frac 1 2} \, . 
\label{eq:case1}
\ee
Case (ii):  $\delta > 0$  (i.e., $1 < \beta$) and  $x < 0$    
\be 
- {\frac N 2 } - {\frac 1 2}  < J_1 \le  - {\frac N 4} - {\frac 1 2} \, . 
\ee
Case (iii):  $\delta < 0 $  (i.e., $0 < \beta < 1$)  and $ x > 0$  \, . 
\be 
{\frac N 4} \le J_1 < {\frac N {\pi} } \tan^{-1}(\sqrt{N-1}) \, . 
\ee
Case (iv):  $\delta < 0$  (i.e., $0 < \beta < 1$)   $ x < 0$   
\be 
- {\frac N {\pi} } \tan^{-1}(\sqrt{N-1}) \le J_1 < - {\frac N 4} \, . 
\ee
Here we recall that the quantum numbers $J_{\alpha}$ are given by 
integers or half-odd integers, which satisfy the condition (\ref{eq:parityJ}).

%%%%%%%%%%%%%%%%%%%%%%%%%%%%%
%
%  Sec 2.3
%
\subsection{Exact number of  missing 2-string solutions}

We now derive  analytically an exact expression for the number of missing 2-string solutions in the XXX spin chain of $N$ sites, $N_{missing}$.  
Here we remark that it is easy to show  the following inequality:  
\be 
 {\frac N {\pi}} \tan^{-1}(\sqrt{N-1}) < {\frac {N-1} 2}  \quad \mbox{for} \, \, N > 2 .  
\ee
We also remark that when $J < (N-1)/2$ the largest Bethe quantum number $J$ satisfying the condition  (\ref{eq:parityJ})  is given by $J=(N-3)/2$ both for even $N$ and odd $N$.

In the case (iii) when $\delta <0$ and $x> 0$ the Bethe quantum number $J_1$ has 
the upper bound $(N/\pi) \tan^{-1}(\sqrt{N-1})$. It corresponds to 
the case of $\beta=0$ (i.e. $\delta=-1/2$), which does not happen for any finite-size system. Therefore,  if the upper bound is equal to the largest possible 
quantum number $(N-3)/2$,  the 2-string solution does not exist. 
Thus,  the number of missing 2-string solutions in the case (iii) is given by  
$[(N-3)/2 - (N/\pi) \tan^{-1}(\sqrt{N-1}) +1]$. Here, the symbol $[x]$ denotes Gauss' symbol, which expresses the largest integer that is less than or equal to $x$.   

By arguing the case (iv) similarly,  
we obtain  an exact estimate $N_{missing}$ 
for the number of missing 2-string solutions 
in the XXX spin chain with $N$ sites as follows. 
\be 
N_{missing} =  2 \left[ \frac  {N-1} 2 - \frac N {\pi} \tan^{-1}(\sqrt{N-1}) \right] \, . 
\label{eq:missing} 
\ee
Here we recall that $[x]$ denotes Gauss' symbol.

%%%%%%%%%%%%%%%%%%%%%%%%%%%%%
%
%  Sec 2.4
%
\subsection{Critical lattice size for  the collapse of a 2-string solution}   

We now derive the critical lattice size $N$ 
for which the number of missing 2-string solutions becomes larger or equal to 1. 
We define  $N_c$ as follows. For $N < N_c$ we have $N_{missing} =0$, 
while for $N \ge N_c$ we have  $N_{missing} \ge 1$, i.e. the 
number of missing 2-string solutions becomes nonzero.  

It was first shown by Essler, Korepin and Schoutens that the critical 
lattice size $N_c$ is given by $N_c=22$ \cite{essler}. 
We confirm it by making use of (\ref{eq:missing})  as follows.  
For $N=22$ we have the largest Bethe quantum number $J_1=(N-3)/2= 
19/2=9.5$. On the other hand,  by putting $N=22$ in  (\ref{eq:missing}) 
we have 
\be 
\frac {22} {\pi} \tan^{-1}(\sqrt{22-1}) =  9.49545\, ,   
\ee
which is slightly smaller than 9.5. Moreover, it is straightforward to show that 
$N_{missing}=0$ for $N < 22$.  
Therefore, the critical lattice size $N_c$ is given by $N_c= 22$.

%%%%%%%%%%%%%%%%%
% Sec 3 
%
\setcounter{equation}{0} 
\renewcommand{\theequation}{3.\arabic{equation}}
\section{Bethe quantum numbers of the real solutions with two-down spins 
}

%%%%%%%%%%%%%%%%%%%%%%%%%%%%%
%
%  Sec 3.1
%
\subsection{Real solutions corresponding collapsed 2-string solutions}  

Let us extend the string deviation $\delta$ to complex values 
by substituting  $\beta = i \gamma$ with  $\gamma \ge 0$ in (\ref{eq:beta}) as follows. 
\be 
\delta = - {\frac 1 2} + {\frac {i} 2} \gamma \, .   \label{eq:def:gamma} 
\ee
We now search for real solutions consisting of two rapidities of the following form:   
\bea 
\lambda_1 & = & x + i (1+ 2 \delta)/2 = x- \gamma \, , \non \\   
\lambda_2 & = & x - i (1+ 2 \delta)/2 = x + \gamma. 
\eea
We call parameter $x$ the center and $\gamma$ the deviation of a real solution with two-down spins, $\lambda_1$ and $\lambda_2$.

By expressing the arctangent functions in terms of the logarithmic functions  
through (\ref{eq:def-arctangent}) with $b=0$  we first show 
\bea
& & 2  \tan^{-1}(2x + i (1+ 2 \delta))  = 2 \tan^{-1} (2x-\gamma) \non \\ 
&  &  =  {\frac 1 {i} } \log \left( \frac {1 + i (2x-\gamma)}  {1 - i (2x-\gamma)} \right)
\non \\
& & =  {\frac 1 {2i}} \log \left( \frac {1 + i (2x-\gamma)}  {1 - i (2x-\gamma)} \right)
  +  {\frac 1 {2i}} \log \left( \frac {1 + i (2x + \gamma)}  {1 - i (2x + \gamma)} 
 \right) \non \\
& & +  {\frac 1 {2i}} \log \left( \frac {1 + i (2x-\gamma)}  {1 - i (2x-\gamma)} \right)
  -  {\frac 1 {2i}} \log \left( \frac {1 + i (2x + \gamma)}  {1 - i (2x + \gamma)} 
 \right) \non \\
&  & = \tan^{-1}(2x+ \gamma) + \tan^{-1}(2x - \gamma) + 
{\frac 1 {2i}} \log \left( \frac {(1- i \gamma)^2 + 4 x^2}
  {(1 +  i \gamma)^2 + 4 x^2}  \right) . 
\label{eq:2x-gamma}
\eea
By making use of  (\ref{eq:2x-gamma}) we have 
\bea 
& & 2 \tan^{-1}(2x + i (1+ 2 \delta)) - {\frac 1 N} 2 \tan^{-1}(i (1+2 \delta))  \non \\ 
& & = \tan^{-1}(2x-\gamma) + \tan^{-1}(2x + \gamma) \non \\  
& & + {\frac 1 {2 i}}  \left\{   
\log \left( \frac {(1-i \gamma)^2 + 4x^2}  {(1+ i \gamma)^2 + 4x^2} \right) 
- {\frac 1 N} \log \left( \frac {(1 - i \gamma)^2}  {(1 + i \gamma)^2} \right) \right\} \, . 
\eea
We now define a counting function $W(x, \gamma)$ by   
\be 
2 \pi W(x, \gamma) =  \tan^{-1}(2 x - \gamma) + \tan^{-1}(2x + \gamma) \, .  
\ee
The Bethe-ansatz equations (\ref{eq:BAE1})
and (\ref{eq:BAE2}) with the complex-valued  string deviation $\delta$
(\ref{eq:def:gamma}) are expressed as follows. 
\bea 
& &  {\frac {2 \pi} N} J_1  =  2 \pi W(x, \gamma)  \non \\ 
& & \quad - \frac 1 {2i} 
\left( \log \left( \frac {(1 + i \gamma)^2 + 4x^2}  {(1- i \gamma)^2 + 4x^2} \right)  
- {\frac 1 N} \log \left( \frac {(1 + i \gamma)^2}  {(1 - i \gamma)^2} \right) \right) \, ,   
\label{eq:BAEcomplex1} \\ 
& &  {\frac {2 \pi} N} J_2  =  2 \pi W(x, \gamma) \non \\
& & \quad + \frac 1 {2i} 
\left( \log \left( \frac {(1 + i \gamma)^2 + 4x^2}  {(1- i \gamma)^2 + 4x^2} \right)  
- {\frac 1 N} \log \left( \frac {(1 + i \gamma)^2}  {(1 - i \gamma)^2} \right) \right)  \, .  
\label{eq:BAEcomplex2}  
\eea 
Here we remark that when we take the $N$th root of  a complex argument of $\log z$, 
we may introduce an $N$th root of unity with an integer $n$ as follows.  
\be 
{\frac 1 N} \log \left( \frac {(1 + i \gamma)^2}  {(1 - i \gamma)^2}  \right) 
=  \log \left\{ \exp \left( {\frac {2\pi i n} N} \right) 
\left( \frac {(1 + i \gamma)^2}  {(1 - i \gamma)^2}  \right)^{1/N} \right\} \, .  
\ee
Suggested by the above remark, 
we now assume the following relation  for $m=0, 1, \ldots, N-1$: 
\be 
 \frac {(1 + i \gamma)^2 +4x^2}  {(1 - i \gamma)^2 +4x^2}  
=  \exp \left( {\frac {2 \pi i m} N } \right) \, \times \,   
\left( \frac {(1 + i \gamma)^2}  {(1 - i \gamma)^2} \right)^{1/N}   \, .    
\label{eq:NthRoot}
\ee 
It follows that the Bethe-ansatz equations 
(\ref{eq:BAEcomplex1}) and (\ref{eq:BAEcomplex2}) are derived from eq. (\ref{eq:NthRoot}) and 
the following equations for $m=0, 1, \ldots, N-1$: 
\bea 
& & {\frac {2 \pi} N} J_1 = 2 \pi W(x, \gamma) \, ,  \label{eq:string-center}  \\ 
& & J_1 =  J_2 - m  \,  . \label{eq:CollapsedRealSol}
\eea

We now solve constraint (\ref{eq:NthRoot}) 
on center $x$ and express $x$  as a function of $\gamma$ with $\gamma \ge 0$. 
Let us introduce variable $\varphi$ by $\varphi= \tan^{-1}\gamma$. 
It satisfies  $0 \le \varphi < \pi/2$.   
From (\ref{eq:NthRoot}) we express the square of center  $x^2$ as a function of $\varphi$ as follows. 
\be 
x^2(\varphi) = \frac 1 {4 \cos^2 \varphi} \frac {\sin \left( 2 \varphi - 
(2\varphi + m \pi)/N \right)} 
{\sin\left( (2 \varphi + m \pi)/N \right)}   \quad (m= 0, 1, \ldots, N-1 ).   
\label{eq:x^2-phi}
\ee
In subsections 3.1 and 3.2 we consider only the case of $m=0$ and 
show that the number of real solutions with $m=0$ is equal to the number of missing complex solutions. However,  the solutions for other values of $m$ correspond to 
the standard real solutions with two 1-strings,  as we shall see in section 3.3.  
We thus derive all the quantum numbers in the two down-spin sector  including both real and complex solutions, simply by extending the string deviation $\delta$ into pure imaginary values.

If we put $m=0$ in (\ref{eq:x^2-phi}), it is clear that the right-hand side is positive   
for the whole range of $\varphi$: $ 0 < \varphi < \pi/2$. 
Thus, by introducing the sign factor: $\epsilon_3= \pm 1$,   
the center $x(\varphi)$ as a function of $\varphi$ is given by 
\be 
x(\varphi) = \frac {\epsilon_3} {2 \cos \varphi} \sqrt{  \frac 
{\sin \left(2 \varphi (1-1/N) \right)}  
{\sin \left( 2 \varphi/N \right)} }  \quad (0 < \varphi < \pi/2)\, . \label{eq:x(varphi)}
\ee 
We shall consider two cases: (v) $x > 0$ ($\epsilon_3=+1$) and (vi) $x< 0$ 
( $\epsilon_3=-1$). 
Through direct calculation we derive  
\be
\frac {d} {d \varphi}  x^2 = \frac 
{\sin(2\varphi/N) \cos (\varphi - 2 \varphi/N) - (1/N) \sin (2 \varphi) 
\cos \varphi }{ 2 \cos^3 \varphi \sin^2(2 \varphi/N) }  \, . \label{eq:dx^2}     
\ee
By proving explicitly that the enumerator of (\ref{eq:dx^2}) is positive 
for $0 < \varphi < \pi/2$  we show that the square of center, $x(\varphi)^2$, 
  is monotonically increasing: 
\be 
 \frac {d } {d \varphi} x^2 > 0 \quad \mbox{for} \, \,  0 < \varphi < \pi/2 \, . 
\label{eq:positivity}
\ee
In order to show the monotonicity of the square of center $x(\varphi)^2$ in shown  (\ref{eq:positivity}) we make use of an inequality: $\cos(\varphi- 2 \varphi/N) > \cos \varphi$ for   $0 <\varphi  < \pi/2$ and the following lemma: 
\begin{lemma} For $\alpha$ satisfying  $0 < \alpha < 1$ we have 
\be 
\sin \alpha x > \alpha \sin x \qquad (0 < x < \pi). 
\label{eq:ineq-alpha}  
\ee
\label{lem:ineq-alpha} 
\end{lemma}
We show lemma \ref{lem:ineq-alpha} by taking the derivative of  
(\ref{eq:ineq-alpha}) with respect to center $x$.     

It thus follows from (\ref{eq:positivity}) that  
we have the minimum value of the square of center 
$x(\varphi)^2$ when we send $\varphi$ to zero. 
\be  
\lim_{\varphi \rightarrow 0} x(\varphi)^2 = {\frac 1 4} {\frac {1- 1/N} {1/N}} = \frac {N-1} 4 \, . 
\ee 
The value of $x(\varphi)^2$ becomes infinite as we send $\varphi$ to $\pi/2$.

We define $\kappa(\varphi)$ by $\kappa(\varphi) = 2 \pi W(x(\varphi), \gamma(\varphi))$   
\be 
\kappa(\varphi) = \tan^{-1}(2 x(\varphi) - \gamma(\varphi) ) 
+ \tan^{-1}(2 x(\varphi)  + \gamma(\varphi) ) \, . \label{eq:kappa}
\ee
We calculate the derivative of $\kappa(\varphi)$ as 
\bea 
\frac {d \kappa} {d \varphi} & = &  
\frac 
 {\displaystyle{4 (1 + 4x^2 + \gamma^2) \frac {dx} {d \varphi}  
- 8 x \gamma {\frac {d \gamma} {d \varphi} }} }
{ \left\{ (2x- \gamma)^2 +1 \right\} \left\{ (2x + \gamma)^2 +1 \right\} }  
\non \\ 
& = &  
\frac 
 { \displaystyle{( 4x^2 + 1/\cos^2 \varphi) \frac {d} {d \varphi} x^2   
- 4 x^2 {\frac {\sin \varphi} {\cos^3 \varphi} } }}
{x/2 \, \left\{ (2x- \gamma)^2 +1 \right\} \left\{ (2x + \gamma)^2 +1 \right\} } \,. 
\label{eq:derivative-kappa}  
\eea  
We shall show in Appendix C the following inequality: 
\be 
\left( 4 x^2 + \frac 1 {\cos^2 \varphi} \right) \frac {d} {d \varphi}  x^2
> 4 x^2 \frac {\sin \varphi} {\cos^3 \varphi}  
\quad ( 0 < \varphi < \pi/2) \, . \label{eq:ineq-dx^2}
\ee
It follows that for  $x > 0$ the derivative of $\kappa(\varphi)$ is positive for $0 < \varphi < \pi/2$ while for  $x < 0$ the derivative of $\kappa(\varphi)$ is negative for $0 < \varphi < \pi/2$. 
Hence, the minimum and the maximum values of $\kappa(\varphi)$ for $x > 0$ are given by 
$\kappa(0)$ and $\kappa(\pi/2)$, respectively, 
while the minimum and the maximum values of $\kappa(\varphi)$ for $x < 0$ 
by $\kappa(\pi/2)$ and $\kappa(0)$, respectively. 

We now consider the case of  $x>0$.   
 We calculate  $\kappa(\pi/2)$ by sending $\varphi$ to $\pi/2$ with $\varphi < \pi/2$. 
In the limit of sending  $\varphi$ to $\pi/2$ 
the term $2x(\varphi) + \gamma(\varphi)$ approaches  
$\infty$ and hence we have 
\be 
\lim_{\varphi \rightarrow \pi/2} \tan^{-1} \left( 2x(\varphi) + \gamma(\varphi) \right) 
= \pi/2 \, .  
\ee
For the term $2x- \gamma$, by putting $\varphi= \pi/2 - \epsilon$ into it and 
by expanding it in terms of $\epsilon$ we show 
\be 
\lim_{\varphi \rightarrow \pi/2}  2x(\varphi) - \gamma(\varphi) 
= \cot \left(\frac {\pi} N \right) \, .  
\ee
We remark that  $\cot x = \tan (\pi/2 -x)$.  We therefore have 
\bea 
\lim_{\varphi \rightarrow \pi/2} \kappa(\varphi)  & = &  
\left( \frac {\pi} 2 - \frac {\pi} N \right) + \frac {\pi} 2  \non \\
& = & \pi - \frac {\pi} N \, . 
\eea
 We calculate  $\kappa(0)$ by sending $\varphi$ to $0$ with $\varphi > 0$. 
\be 
\lim_{\varphi \rightarrow 0} \kappa(\varphi) = 2 \tan^{-1} \left( \sqrt{N-1}  \right) .
\ee

We thus have the following results. 
\par \noindent 
Case (v) : $x > 0$ with $\delta=(-1+ i \gamma)/2$   
\be 
\frac N {\pi} \tan^{-1}(\sqrt{N-1}) \le J_1 < \frac N 2 - \frac 1 2   
\ee
Case (vi) : $x < 0$ with $\delta=(-1+ i \gamma)/2$   
\be 
 - \frac N 2 + \frac 1 2  < J_1 \le  - \frac N {\pi} \tan^{-1}(\sqrt{N-1})
\ee
Here we assume that if $\varphi=0$ we have a real solution with a pair of 
the same rapidities.

%%%%%%%%%%%%%%%%%%%%%%%%%%%%%
%
%  Sec 3.2
%
\subsection{Number of new real solutions corresponds 
to that of missing 2-string solutions}  

In the case (v) when $x > 0$ with complex-valued 
 deviation $\delta$, the largest Bethe quantum number $J_1$ is given by 
$(N-3)/2$ both for even and odd $N$, as shown in subsection 2.3. 
Here the smallest Bethe quantum number is given by 
an integer or a half-integer greater than or equal to 
$(N/\pi) \tan^{-1} \left( \sqrt{N-1} \right)$. Therefore,  
the number of new real solutions in the case (v) 
is given by  $N_{new} = [(N-3)/2 - N/\pi \tan^{-1} \left( \sqrt{N-1} \right) + 1 ]$. 

Similarly, we have in the case (vi) when $x< 0$, we have
the number of  new real solutions 
$ [(N-3)/2 - N/\pi \tan^{-1} \left( \sqrt{N-1} \right) + 1 ]$. 
In total, we have the same number in the case of (vi).

Combining two cases (v) and (vi) we give the number of 
new real solutions is given by   
\be 
N_{new} =  2 \left[ \frac  {N-1} 2 - \frac N {\pi} \tan^{-1}(\sqrt{N-1}) \right] \, . 
\label{eq:new} 
\ee
Thus, the number of missing complex solutions is exactly equal to the number of 
new real solutions. Moreover, the new real solutions 
have the same Bethe quantum numbers with those of the missing complex solutions.   

%%%%%%%%%%%%%%%%%%%%%%%%%%%%%
%
%  Sec 3.3
%
\subsection{Standard real solutions as 2-strings with imaginary deviations}  

In the case of $m \ne 0$ such as $m=1, 2, \ldots, N-1$  for eq. (\ref{eq:x^2-phi}) 
 we express $x(\varphi)$ as follows. 
\be 
x(\varphi) = \frac {\epsilon_3}  {2 \cos \varphi} \sqrt{ \frac {\sin \left( 2 \varphi - 
(2\varphi + m \pi)/N \right)} 
{\sin\left( (2 \varphi + m \pi)/N \right)} }  \, . 
\label{eq:x^2-phi-m}
\ee
Since $x(\varphi)^2$ must be non-negative, the range of parameter $\varphi$ is given  by 
\be 
{\frac {m \pi} {2(N-1)}} \le \varphi <  \frac {\pi} 2 \quad  
\mbox{for} \quad m=1, 2, \ldots, N-1.  \label{eq:int-m}
\ee
Here we do not consider the case of $m=N-1$ since there is no range for  
$\varphi$ satisfying (\ref{eq:int-m}).   
We denote the minimum value of $\varphi$ by $\varphi_{min}$: 
$\varphi= m \pi/2(N-1)$. 
By taking  the derivative of the square of center $x(\varphi)^2$ with respect to 
variable $\varphi$, we have  
\bea
& &\frac {d} {d \varphi}  x(\varphi)^2  
\non \\ 
& & =  
\frac {\sin \left( (2\varphi + m \pi)/N \right) \cos (\varphi - (2 \varphi + m \pi)/N) 
- (1/N) \sin (2 \varphi) \cos \varphi }
{ 2 \cos^3 \varphi \sin^2((2 \varphi+m \pi)/N) }  \, . \non \\ 
\label{eq:dx^2-m}     
\eea
Proving explicitly that the enumerator of (\ref{eq:dx^2-m}) is positive 
for $m \pi/(2 (N-1)) \le \varphi < \pi/2$ in Appendix C,  
we show that the square of center, $x(\varphi)^2$, 
is monotonically increasing in the region: $\varphi_{min} \le \varphi < \pi/2$: 
\be 
 \frac {d } {d \varphi} x^2 > 0 \qquad \mbox{for} \quad  
{\frac {m \pi} {2(N-1)}} < \varphi < \pi/2 \, . 
\label{eq:positivity-m}
\ee
In order to show (\ref{eq:positivity-m}) 
we make use of an inequality: 
\be 
\cos \left(\varphi- {\frac {2 \varphi + m \pi} N} \right) > \cos \varphi  \quad 
\mbox{for } 
\quad {\frac {m \pi}  {2 (N-1)}} < \varphi < {\frac {\pi} 2} . 
\label{eq:inq-cos-m}
\ee
and the next lemma.  
\begin{lemma} For any given real number $N$ satisfying $N > 1$  we have 
\be 
\sin  \left( \frac {2(\varphi+ m \pi)} N \right)  > {\frac 1 N} \sin 2 \varphi \quad \mbox{for} \quad  
{\frac {m \pi}  {2 (N-1)}} < \varphi < {\frac {\pi} 2} . 
\label{eq:ineq-alpha-m}  
\ee
\label{lem:ineq-alpha-m} 
\end{lemma}
We can prove lemma \ref{lem:ineq-alpha-m} by taking the derivative of 
(\ref{eq:ineq-alpha-m}) with respect to center $x$.   

It follows from (\ref{eq:inq-cos-m}) and (\ref{eq:ineq-alpha-m})  
that we have the following inequality: 
\bea 
& & \sin  \left( \frac {2(\varphi+ m \pi)} N \right) 
\cos \left(\varphi- {\frac {2 \varphi + m \pi} N} \right) > {\frac 1 N} \sin 2 \varphi \cos \varphi  
\non \\ 
& & \qquad \qquad \mbox{for} \quad  
{\frac {m \pi}  {2 (N-1)}} < \varphi < {\frac {\pi} 2} . 
\eea
It follows that inequality (\ref{eq:positivity-m}) holds, and hence the square of center 
$x(\varphi)^2$ is monotonic increasing.  

We define function $\kappa(\varphi)$ by (\ref{eq:kappa}) also for the cases of 
$m=1, 2, \ldots, N-1$. Here we recall that  
the derivative of $\kappa(\varphi)$ with respect 
to $\varphi$ is given in eq.  (\ref{eq:derivative-kappa}).   
We shall show in Appendix D the following inequality for $m=1, 2, \ldots, N-2$:  
\be 
\left( 4 x^2 + \frac 1 {\cos^2 \varphi} \right) \frac {d} {d \varphi}  x^2
> 4 x^2 \frac {\sin \varphi} {\cos^3 \varphi}  
\quad \mbox{for} \quad 
{\frac {m \pi} {2(N-1)}} \le \varphi <  \frac {\pi} 2 \,  . \label{eq:ineq-dx^2-m}
\ee
It therefore follows that $\kappa(\varphi)$ is monotonically increasing function 
in the interval with  $\pi/(2 (N-1)) < \varphi < \pi /2$.

%%%%%%%%%

We consider case (vii) when   $x > 0$ and case (viii) when  $x < 0$. 
For  $x > 0$ we can show that the derivative of $\kappa(\varphi)$ is positive for 
$\varphi_{min} < \varphi < \pi/2$. 
Hence, the minimum and maximum of $\kappa(\varphi)$ are given by 
$\kappa(\varphi_{min})$ and $\kappa(\pi/2)$, respectively. 
We calculate  $\kappa(\pi/2)$ by sending $\varphi$ 
to $\pi/2$ with $\varphi < \pi/2$. 
In the limit of sending  $\varphi$ to $\pi/2$ 
the term $2x(\varphi) + \gamma(\varphi)$ approaches  infinity, 
and hence we have 
\be 
\lim_{\varphi \rightarrow \pi/2} \tan^{-1} \left( 2x(\varphi) + \gamma(\varphi) \right) 
= \pi/2 \, , 
\ee
For $2x- \gamma$, by putting $\varphi=\pi/2 - \epsilon$ and expanding 
(\ref{eq:x^2-phi-m}) with respect to $\epsilon$, we show 
\be 
\lim_{\varphi \rightarrow \pi/2}  2x(\varphi) - \gamma(\varphi) 
= \cot \left(\frac {(m+1) \pi} N \right) \, .  
\ee
Hence we have 
\be 
\lim_{\varphi \rightarrow \pi/2} \kappa(\varphi)   =   
{\pi} - \frac {(m+1) \pi} N  \, .   
\ee
 We calculate  $\kappa(\varphi_{min})$ by sending $\varphi$ to $\varphi_{min}$ 
with $\varphi >  \varphi_{min}$. 
\be 
\lim_{\varphi \rightarrow \varphi_{min}} \kappa(\varphi) = 0 \, . 
\ee

We therefore have the following results. 
\par \noindent 
Case (vii):  $x > 0$ with complex deviation
\be 
- {\frac m 2}  \le J_1 < {\frac N 2} - {\frac 1 2} - m \, .      \label{case7}
\ee
Case (viii): $x < 0$  with complex deviation
\be 
 - \frac N 2 + \frac 1 2  < J_1 \le  - {\frac m 2} \, .       \label{case8}
\ee

Combining case (vii) and (viii) we have 
\be 
- {\frac N 2} + {\frac 1 2}  < J_1 <  {\frac N 2} - {\frac 1 2} - m \, . \label{eq:real}
\ee
Here we recall that $J_2$ is given by eq. (\ref{eq:CollapsedRealSol}): 
$J_2 = J_1 +m$ for $m=1, 2, \ldots, N-1$. 
It is easy to show that the set of the Bethe quantum numbers $J_1$ and $J_2$ satisfying (\ref{eq:real}) corresponds to the set  of $J_1$ and $J_2$ satisfying the following conditions: 
\be 
-\frac {N-2} 2 < J_1 < J_2 < \frac {N-2} 2 \, . 
\ee
They are nothing but the conditions of the quantum numbers for the standard 1-string solutions  with $M=2$ \cite{taka,takab}. Here we remark that when $N$ is even $J_j$s are half-integers and hence $J_j \le (N-2)/2 -1/2 = (N-3)/2$, which is equivalent to $J_j < (N-1)/2$.    

It is easy to show that the number of  pairs $J_1$ and $J_2$ satisfying  (\ref{eq:CollapsedRealSol}) and (\ref{eq:real})  is given by $(N-2)(N-3)/2$, which 
coincides with the number of 1-strings in the $M=2$ sector expected by the string hypothesis.  Thus, by deriving all the Bethe quantum numbers $J_1$ and $J_2$ 
exactly, we have shown the number of standard real solutions 
in the two down-spin sector. Moreover,  it is consistent with the string hypothesis.

%%%%%%%%%%%%%%%%%%%%%%%%%%%%%%
%  Sec 4 
%
\setcounter{equation}{0} 
\renewcommand{\theequation}{4.\arabic{equation}}
\section{Some features of 2-string solutions}

%%%%%%%%%%%%%%%%%%%%%%%%%%%%%
%
%  Sec 4.1
%
\subsection{Singular string solution}

We now show that when $N$ is given by $N=4n$ with an integer $n$, 
the Bethe quantum numbers $(J_1, J_2)  = (N/4 - 1/2, N/4 + 1/2)$ and 
$ (-N/4 - 1/2, - N/4 + 1/2) $, which correspond to  
the cases (i) and (ii), respectively, give the singular string solution   
$(\lambda_1, \lambda_2) =(i/2, -i/2)$. 
We also show that when $N$ is given by $N=4n+2$ with an integer $n$, 
the Bethe quantum numbers $(J_1, J_2)  = (N/4, N/4)$ and 
$ (-N/4, - N/4) $, which correspond to the cases (iii) and (iv), respectively,  
 give the singular string solution   $(\lambda_1, \lambda_2) =(i/2, -i/2)$.

For  $N=4n$ we have  $N/4 - 1/2= n -1/2$, and it is  a half-integer. 
Since $N$ and $M$ are even, the Bethe quantum numbers $J_1$ and $J_2$  
are given by half-integers. Thus, $J_1 = N/4 - 1/2$ (mod 1), and hence $J_1$ can take the 
value $N/4 - 1/2$.  In the limit of sending the string deviation 
$\delta$ to 0 with $\delta > 0$ we have  
\bea 
& &  \lim_{w \rightarrow 0; w> 0}  N Z(x(w), \delta(w)) \non \\ 
& = & \lim_{w \rightarrow 0; w> 0} 
\frac N {2 \pi}\left( \pi H(\delta) \left({\rm sgn}(x) - \frac 1 N \right) 
+ \tan^{-1} \left( \frac x {1+ \delta}(w)  \right) -  \tan^{-1} \left(  {\frac x {\delta}}(w)  \right) \right) \non \\ 
& = & { \frac  N {2 \pi}} \left( \pi (1- {\frac 1 N}) 
+ \tan^{-1}( 0) - \tan^{-1}( \infty) \right)   
\non \\ 
& = & {\frac N 4} - \frac 1 2 \, . 
\eea
Here we have made use of the small $w$-behavior:
\be 
{ x /{(1 + \delta)}} \approx w ^{1/2N} \, \,  (w \ll 1),  
\qquad 
{ x /{\delta}} \approx w^{1/2N - 1/2} \, \,  (w \ll 1).  
\ee
We have $J_2 = J_1 +1$ since $\delta > 0$. We have 
$J_2 = J_1 + 1 = N/4 + 1/2$. 
Thus, we have derived the singular solution 
by sending the string deviation $\delta$ to 0 with $\delta > 0$. 
By setting $\epsilon= w^{1/2N}$, we have 
\be 
\lambda_1 = \epsilon + i/2 + i \epsilon^N \, ,  \qquad 
\lambda_2 = \epsilon - i/2 - i \epsilon^N \, . 
\ee
It is nothing but one of  the regularization schemes for the singular string solution, 
where  we send the positive small parameter $\epsilon$ to zero.  

In the case of $(J_1, J_2)  = (-N/4 - 1/2, -N/4 + 1/2)$,  
we consider case (ii) $\delta > 0$ and $x < 0$. 
We derive $J_1= -N/4 -1/2$ from the following limit:  
\bea 
& & \lim_{w \rightarrow 0; w> 0} 
{\frac N {2 \pi }} \left( \pi H(\delta) \left({\rm sgn}(x) - \frac 1 N \right) 
+ \tan^{-1} \left( \frac x {1+ \delta}(w)  \right) 
-  \tan^{-1} \left(  {\frac x {\delta}}(w)  \right) \right) \non \\ 
& = & { \frac  N {2 \pi}}  \left( \pi ( -1- {\frac 1 N}) + \tan^{-1}( 0) - \tan^{-1}( - \infty) \right)   
\non \\ 
& = & - {\frac N 4} - \frac 1 2 \, . 
\eea
We thus obtain $J_1= -N/4 -1/2$. 
By setting $\epsilon= w^{1/2N}$ we have 
\be 
\lambda_1 = - \epsilon + i/2 + i \epsilon^N \, ,  \qquad 
\lambda_2 = - \epsilon - i/2 - i \epsilon^N \, . 
\ee
Here, the string center $x$  is negative:  $x= - \epsilon$ and we send 
the small positive parameter $\epsilon$ to zero.

Similarly, we can show that for  $N=4n+2$ with an integer $n$, 
the Bethe quantum numbers $(J_1, J_2)  = (N/4, N/4)$ and $ (-N/4, - N/4)$ 
give the singular string solution: $(\lambda_1, \lambda_2) =(i/2, -i/2)$,  
by considering cases (iii) $\delta < 0$ and $x> 0$ and (iv) $\delta < 0$ and $x< 0$, respectively. 
In the case of  $(J_1, J_2)  = (N/4, N/4)$, 
by setting $\epsilon= w^{1/2N}$ we have 
\be 
\lambda_1 =  \epsilon + i/2 - i \epsilon^N \, ,  \qquad 
\lambda_2 = \epsilon - i/2 + i \epsilon^N \, . 
\ee
Here, the string center $x$ is positive:  $x= \epsilon$, 
while the string deviation $\delta$ is negative: $\delta= - \epsilon^N$, 
and we send the small positive parameter $\epsilon$ to zero.

In the case of  $(J_1, J_2)  = (-N/4, -N/4)$,  
by setting $\epsilon= w^{1/2N}$ we have 
\be 
\lambda_1 = - \epsilon + i/2 - i \epsilon^N \, ,  \qquad 
\lambda_2 = - \epsilon - i/2 + i \epsilon^N \, . 
\ee
Here, the string center $x$ is negative:  $x= - \epsilon$, 
and the string deviation $\delta$ is negative: $\delta= - \epsilon^N$,    
and we send them to zero. 

Here we remark that it has been observed through numerical solutions that 
the different quantum numbers correspond to the same singular string solution \cite{fuji,hag}. 
However,  we have derived the four different sets of the Bethe quantum numbers corresponding to the same singular string solution,  
by solving the Bethe-ansatz equations in the logarithmic form 
with an analytic approach of sending the string deviation $\delta$ to zero. 
We recall that it is not possible to put $\delta=0$  directly in the BAEs.   
Hence, the singular string solution is different 
from the standard generic solutions of the BAEs. It satisfies the BAEs only in the limiting procedure. 

%%%%%%%%%%%%%%%%%%%%%%%%%%%%%%%%%%%%%%%%%%%%%
%
\subsection{Large-$N$ behavior of  2-string solutions}

Let us consider case  (i) when $ \delta > 0$ and $x> 0$. 
We first recall  (\ref{eq:ratios}). Sending $w$ to 1 we have 
\be 
\lim_{w \rightarrow 1}  \frac x {\delta} = \sqrt{N-1} \, . 
\ee 
When $w$ is close to 1,  both $x$ and $\delta$ are very large and  
we have $x \approx \sqrt{N-1} \delta$.  
Secondly, the largest Bethe quantum number $J_1$ satisfying (\ref{eq:case1})
is given by $(N-3)/2$. 
Putting $J_1=(N-3)/2$ in to  (\ref{BAE1Re}) we have 
\be 
- {\frac {2\pi} N} = \tan^{-1}(x/(1+ \delta)) - \tan^{-1}(x/\delta) \, .    
\ee
Assuming that $x$ is large we apply the expansion: 
$\tan^{-1}(x) = \pi/2 - 1/x +  \cdots$, and we evaluate $x$ as $x \approx  N/2 \pi$. 
Thus, for $J_1=(N-3)/2$, we have 
\be 
\delta \approx \frac {\sqrt{N}} {2 \pi} \, .  
\ee

%%%%%%%%%%%%%%%%%%%%%%%%%%%%%%
%  Sec 5 
%
\setcounter{equation}{0} 
\renewcommand{\theequation}{5.\arabic{equation}}
\section{Completeness of the Bethe ansatz in the two down-spin sector} 

\subsection{Enumeration of collapsed and non-collapsed 2-string solutions}

Let us now enumerate the number of solutions in the cases (i) to (vi). 
We combine the case (iii) with the case (v) and the case (iv) with the case (vi).

\par \noindent 
(A): $N=4n$ 

We have one singular solution for either the case (i) or (ii), and $n-1$ generic solutions in each of the four cases, i.e., the cases (i), (ii),  (iii) $\&$ (v), and (iv) $\&$ (vi).  
In total, we have  $N-3$ solutions,  since $4(n-1) +1 = 4n- 3 = N-3$. 

\par \noindent 
(B): $N=4n+1$ 

We have no singular solution, and $n$ generic solutions in the cases (i) and (ii), 
$n-1$ generic solutions in the cases (iii) $\&$ (v) and (iv) $\&$ (vi).  
In total, we have  $N-3$ solutions,  since $2 n + 2 (n-1) = 4n- 2 = N-3$.

\par \noindent 
(C): $N=4n+2$ 

We have one singular solution in either the case (iii) or (iv),  $n$ generic solutions in  the cases (i) and (ii), and $n-1$ generic solutions in the cases (iii) and (iv).
In total, we have  $N-3$ solutions,  since $2n+ 2(n-1) +1 = 4n- 1 = N-3$.

\par \noindent 
(D): $N=4n+3$   

We have no singular solution, and $n$ generic solutions in each of the four cases.  
In total, we have  $N-3$ solutions,  since $2 n + 2 n = 4n  = N-3$. 

Thus, we have shown that there are  $(N-3)$ 2-string solutions for any number of the lattice size $N$.

%%%%%%%%%%%%%%%%%%%%%%%%%%%%%%%%%%%%
\subsection{Completeness through enumeration of the Bethe quantum numbers}
We have $(N-2)(N-3)/2$ solutions for the 1-string and $N-3$ solutions for the 2-string. 
In total we have 
\bea 
(N-2)(N-3)/2 + N-3 & = & N(N-3)/2 \non \\ 
& = & N(N-1)/2 -N \non \\ 
& = & _NC_2 - _NC_1 
\eea   
The number is consistent with that of the string hypothesis, which is given by  
 the number of highest weight vectors  under the total spin $SU(2)$ symmetry.

\subsection{Analytic derivation of the Takahashi quantum numbers}

Let us assume a 2-string solution $\lambda_{1}= x + i(1 + 2 \delta)/2$ and  
$\lambda_{2}= x - i (1 + 2 \delta)/2$.  
We define the Takahashi quantum number $I$ for the solution by 
\be 
2 \tan^{-1}\left( \frac {x} {1 + \delta} \right) = \frac {2 \pi} N I   \, . 
\label{eq:def-I}
\ee
We can show 
\be 
\lim_{w \rightarrow 0} 2 \tan^{-1} \left( \frac {x(w)} {1 + \delta(w)} \right)
= {2 \pi} N (J_1 + J_2) - \pi {\rm sgn}( x ) \, . \label{eq:limit}
\ee
Therefore, we have the relation between the Takahashi quantum numebers $I$ and 
Bethe quantum numbers $J_1$ and $J_2$ as follows. 
\be 
I = J_1 + J_2 - {\frac N 2 } {\rm sgn} (x) \, . \label{eq:rel}
\ee

Let us now explain the derivation of the limit (\ref{eq:limit}). 
For $2 \tan^{-1} 2 \lambda_1$ we recall eq. (\ref{eq:x-d}).  
For $2 \tan^{-1} 2 \lambda_2$ we have the following. 
\bea 
& & 2 \tan^{-1} 2 \lambda_2   = 2 \tan^{-1}(2x - i (1+ 2\delta)) \non \\ 
& &  =  \tan^{-1} \left( \frac x {1+ \delta}  \right) -  \tan^{-1} \left(  \frac x {\delta} \right)  \non \\ 
& & \qquad + \pi H(\delta) {\rm sgn}(x_{+})  +  
\frac 1 {2 i} \log \left( \frac {x^2 + \delta^2} {x^2 + (1+ \delta)^2}   \right) \, .  
\label{eq:x-dd}
\eea 
We therefore have the following: 
\bea 
& & 2 \tan^{-1} 2 \lambda_1 + 2 \tan^{-2} 2 \lambda_2   \non \\ 
& & =  2 \tan^{-1}(2x + i (1+ 2\delta)) +  2 \tan^{-1}(2x - i (1+ 2\delta))
\non \\ 
& & =  2 \tan^{-1} \left( \frac x {1+ \delta}  \right) -  2 \tan^{-1} \left(  \frac x {\delta} \right)  + \pi H(\delta) {\rm sgn}( (x_{+}) +  {\rm sgn}(x_{-})) \, . 
\label{eq:sumAB}
\eea
Here we remark that when $\delta$ is very small, then $w$ is also very small since 
we have eq. (\ref{eq:ratios}). When we take the limit of sending $w$ to zero, 
we have 
\be 
\frac x {\delta} = \epsilon_2 \sqrt{\frac {w^{1/N-1}-1} {1 -w^{1/N}}}    
\rightarrow  \epsilon_2 w^{1/2N-1/2} \quad ( w \rightarrow 0) . 
\ee
Hence we have 
\be 
2 \tan^{-1}  \left(  \frac x {\delta} \right)
\rightarrow 2 \tan^{-1} \left( \epsilon_2 w^{1/2N - 1/2}  \right) = \epsilon_2 \pi 
\quad (w \rightarrow 0). 
\ee
Here we recall the relation: $\epsilon_2 = {\rm sgn}(\delta) {\rm sgn}( x)$. 
By taking the sum of eqs. (\ref{eq:BAE1}) and (\ref{eq:BAE2}) we have 
\be 
2 \tan^{-1} 2 \lambda_1 + 2 \tan^{-2} 2 \lambda_2 = \frac {2 \pi} N (J_1 + J_2) \, . 
\label{eq:sum}
\ee
It follows from eq. (\ref{eq:sumAB}). that the left-hand side of eq. (\ref{eq:sum}) is given by 
\bea 
& & 2 \tan^{-1} 2 \lambda_1 + 2 \tan^{-2} 2 \lambda_2 
\non \\ 
& & =  2 \tan^{-1} \left( \frac x {1+ \delta}  \right) -  2 \tan^{-1} \left(  \frac x {\delta} \right)  + \pi H(\delta) {\rm sgn}( (x_{+}) +  {\rm sgn}(x_{-})) \, .
\non \\ 
& &  = 2 \tan^{-1} \left( \frac x {1+ \delta}  \right) - \pi {\rm sgn}(\delta) 
{\rm sgn} (x)   + \pi H(\delta)( {\rm sgn}(x_{+}) +  {\rm sgn}(x_{-}) ) 
\non \\ 
& &  = 2 \tan^{-1} \left( \frac x {1+ \delta}  \right) + \pi \, {\rm sgn}(x) \, .  
\eea
Here we note 
\bea 
& & - \pi {\rm sgn}(\delta) {\rm sgn} (x)   + \pi H(\delta) ({\rm sgn} (x_{+}) +  {\rm sgn}(x_{-})) 
\non \\ 
& & = \left\{
\begin{array}{cc} 
- \pi {\rm sgn} (x)   + 2 \pi {\rm sgn}(x) =  \pi {\rm sgn} (x)  & \delta > 0 \, ,  \\ 
 \pi {\rm sgn} (x)    & \delta < 0  \, . \\
\end{array} \right.
\eea
Through the definition (\ref{eq:def-I}) we obtain the relation (\ref{eq:rel}) 
between the Bethe quantum numbers $J_1$ and $J_2$ and 
the Takahashi quantum number $I$. 

Considering the four cases:  (A) $N=4n$, (B) $N=4n+1$, (C) $N=4n+2$ and (D) $N=4n+3$,  
we can show that the Takahashi numbers $I$ of the non-collapsed and collapsed 2-string solutions are distinct for all the $N-3$ solutions in any system size $N$.   
They are given by $-N/2 + 2, - N/2 + 3, \ldots,  N/2-2$ for any integer $N$.

%%%%%%%%%%%%%%%%%%%%%%%%%%%%%%
%  Sec 6 
%
\setcounter{equation}{0} 
\renewcommand{\theequation}{6.\arabic{equation}}
\section{Discussion and concluding remarks} 
\subsection{On the proof of the completeness}

In the paper we have derived the Bethe quantum numbers rigorously for 
all the solutions of the XXX spin chain in the two down-spin  sector. 

Since any complex solution with two down-spins is expressed in terms of two real parameters we have first solved BAE with respect to one of the two parameters and introduced the counting functions which give quantum numbers as their special values. 
We have shown that they are monotonic by explicitly calculating their derivatives and given analytically the upper and lower bounds to them.  We have thus obtained the Bethe quantum numbers both for real and complex solutions in the two down-spin sector  without making any assumption.  We have thus shown that there are no more quantum numbers for physical solutions of BAE. We remark that even for real solutions we have sown that no other  Bethe quantum numbers exist. 

Moreover, we have shown that the derived set of the Bethe quantum numbers 
give exactly the same number to the dimensions of the vector space spanned by the eigenvectors of the XXX spin chain in the two down-spin sector. 
Therefore, if the Bethe-ansatz eigenvectors of the complex and real solutions 
associated with the derived set of the Bethe quantum numbers  are linearly independent, 
then the derived Bethe-ansatz eigenvectors give a complete set of the subspace in the two down-spin sector.  

It is shown by Slavnov that the inner product of two different Bethe-ansatz eigenvectors vanishes \cite{Slavnov}. Therefore, if we assume that the solutions of BAE for the derived set of the Bethe quantum numbers are distinct, then they lead to a complete set of the eigenvectors in the two down-spin sector.    
Although mathematically it has not been shown, yet, but it is quite likely that 
the solutions of BAE for the derived set of the Bethe quantum numbers are distinct, 
since the derived set consists of distinct pairs of integers or half-integers. 
   
\subsection{Suggestions for further studies}

In the case of three down-spins the parametrization introduced in the paper is still useful 
and we can show some properties of the Bethe quantum numbers such as 
the difference of the quantum numbers associated with the string deviations. 
Some details will be given elsewhere.  

%\vskip 24pt Extension to the XXZ spin chain 

Moreover, we can calculate the Bethe quantum numbers for 
the XXZ spin chain in the two down-spin sector by only slightly extending the method 
in the paper.  For instance, we can derive the quantum numbers of  2-string solutions 
for the spin-1/2  XXZ chain in the gapful anti-ferromagnetic regime.  
We consider the XXZ Hamiltonian under the periodic boundary conditions: 
\be  
{\cal H}_{\rm XXZ} = {\frac 1 2} \sum_{j=1}^{N} \left( \sigma_{j}^{x} \sigma_{j+1}^{x}
+ \sigma_{j}^{y} \sigma_{j+1}^{y}
+ \Delta \sigma_{j}^{z} \sigma_{j+1}^{z} \right) \,  , 
\ee
where $\sigma_j^a$ ($a=x,y,z$) are the Pauli matrices defined on the $j$th lattice site.  
Let us express the XXZ anisotropy $\Delta$ by 
$\Delta = \cosh \zeta$ with  $\zeta>0$. 
The Bethe-ansatz equations are given by 
\bea 
2 \tan^{-1}\left( {\frac {\tan \lambda_j} {\tanh (\zeta/2)}} \right) 
& = & {\frac {2 \pi} N} J_j + {\frac 1 N} \sum_{k=1}^{M}  2 \tan^{-1} \left( 
{\frac {\tan(\lambda_j - \lambda_k)} {\tanh \zeta}} \right) \non \\ 
& & \mbox{ for} \quad j = 1, 2, \ldots, M . 
\eea
Here the quantum numbers $J_j$ satisfy the condition: 
$J_j = (N-M+1)/2$ (mod 1).  We assume the folowing form of a 2-string: 
\be 
\lambda_1=  x + {\frac i 2} \zeta + i \delta, \quad 
\lambda_2=  x - {\frac i 2} \zeta - i \delta \, . 
\ee
We can express the string center $x$ in terms of deformation parameter 
$w= \tanh(\zeta/2 + \delta)/\tanh (\zeta/2)$ and $z= \tanh(\zeta/2)$. 
Explicitly, the string center $x$ is given by 
\be 
\tan^2 x = {\frac 1 { 2 {\cal A}}} 
\left( - {\cal B} \pm \sqrt{{\cal B}^2 - 4 {\cal A} {\cal C} } \right) 
\ee
where ${\cal A}$, ${\cal B}$ and ${\cal C}$ are given by 
\bea 
{\cal A} & = & w^2(1+w z^2)^2 \left({\frac {1-w} {1+w}} {\frac {1-w z^2} {1 + w z^2}} \right)^{2/N} - w^2 (1- w z^2)^2 \, , \non \\  
{\cal B} & = & \left\{ (1 + w^2 z^2)^2/z^2 + 2w (1+w)(1+w z^2) \right\} 
 \left({\frac {1-w} {1+w}} {\frac {1-w z^2} {1 + w z^2}} \right)^{2/N}  \non \\ 
& & \qquad - \left\{ (1 + w^2 z^2)^2/z^2 - 2w (1-w)(1-w z^2) \right\} \, , \non \\  
{\cal C} & = & (1+w)^2  \left({\frac {1-w} {1+w}} {\frac {1-w z^2} {1 + w z^2}} \right)^{2/N} - (1-w)^2 \, .
\eea 
We show that string deviation $\delta$ is negative with its absolute value being very small in order to make the string center $x$ being real-valued. 
Some more details will be discussed elsewhere.

%%%%%%%%%%%%%%%%%%%%%%%%%%%%%%%%%%%%%%%%%%
\appendix 
\setcounter{equation}{0} 
\renewcommand{\theequation}{A.\arabic{equation}}
\section{Formulas related to the logarithmic function}

For a given nonzero complex number 
$z = \alpha + i \beta$  where $\alpha$ and $\beta$ are real  
we express the logarithmic function $\log z$ by   
\be 
\log\left( \alpha + i \beta \right) = i \left( \theta(z) + 2 \pi n \right) 
 + {\frac 1 {2}} \log( \alpha^2 + \beta^2)   \, ,   
\label{eq:log-z}
\ee
where $n$ is an integer ($n \in {\bm Z}$) corresponding to the branch 
of the logarithmic function and we express  $\theta(z)$ as   
\be
\theta(\alpha + i \beta) = \left\{ \begin{array}{cc}  
\tan^{-1}\left({\beta}/{\alpha} \right)
+ \pi H(- \alpha) {\rm sgn}(\beta_{+}) & \mbox{for}  \, \, \alpha \ne 0 \\ 
  {\rm sgn}(\beta_{+}) \, {\pi}/2 &  \mbox{for}  \, \,  \alpha=0    
\end{array} \right. 
\label{eq:theta}
\ee
Here we recall  that 
we take the branch: $-\pi/2  < \tan^{-1} x < \pi/2$ for $(x \in {\bm R})$.  
We denote by  ${\rm sgn}(x_{+})$ a sign function ${\rm sgn}(x)$ 
shifted by an infinitesimally small positive number $0_{+}$: 
${\rm sgn}(x_{+}) = {\rm sgn}(x + 0_+) = 1 - 2 H(-x)$.

The function $\theta(\alpha + i \beta)$ defined by (\ref{eq:theta}) 
is continuous at $\alpha=0$ as a function of $\alpha$ when  $\beta \ne 0$. 
The range is given by $- \pi < \theta(z) < \pi$ if  $\beta \ne 0$, 
while if $\beta=0$  $\theta(z) =0$ or  $\pi$ for $\alpha > 0$ or $\alpha < 0$, respectively.

We define the arctangent function $\tan^{-1}(a+ i b)$  
for a nonzero complex number $a+ i b$ where $a$ and $b$ are real by 
\be 
  \tan^{-1}(a+ i b)  = {\frac 1 {2i} } \left( \log(1-b +i a) - \log(1+b - i a) \right)
\label{eq:def-arctangent}
\ee

Applying formula (\ref{eq:log-z}) to (\ref{eq:def-arctangent}) 
we have for $b \ne \pm1$ 
\bea
2 \tan^{-1}(a+ ib)  
& & = \tan^{-1}\left( \frac a {1-b} \right) + \pi H(b-1) {\rm sgn}(a_{+}) 
\non \\ 
& & +  \tan^{-1}\left( \frac a {1+b} \right) +   \pi H(-b-1) {\rm sgn}(a_{-}) \non \\ 
& & + {\frac 1 {2i}} \log \left( \frac {a^2 + (b-1)^2} {a^2 + (b+1)^2 } \right)  \, . 
\label{eq:arctan}
\eea

In the branch: $-\pi < {\rm Im} \log z \le \pi$ we can show   
\bea 
 - \pi < & & {\rm Re}\left(  2 \tan^{-1}(a+ ib) \right) < \pi  \quad ( a \ne 0 ),    
\non \\ 
& &  {\rm Re}\left(  2 \tan^{-1}(a+ ib) \right) = \pm \pi  \quad ( a = 0 ).   
\eea

%%%%%%%%%%%%%%%%%%%%%%%%%%%%%%%%%%%%%%%%%%%%
\setcounter{equation}{0} 
\renewcommand{\theequation}{B.\arabic{equation}}
\section{Proof of lemma \ref{lem:2} }

By dividing each side of  inequality (\ref{eq:ineq2}) by $(1+ w ^{1/2})$ 
we reduce it to the following: 
\be 
(1- w^{1/2})(1 + w^{1/N-1/2}) > N(1-w^{1/N})    \, . 
\ee
Let us define $F(w)$ by 
$F(w) = (1- w^{1/2})(1 + w^{1/N-1/2}) - N(1-w^{1/N})$. 
By taking the derivative  we have  
\be 
F^{'}(w) = - \frac {w^{1/N- 3/2}} 2
 \left( w^{1-1/N} - 2(1- {\frac 1 N}) w^{1/2} + 1 - {\frac 2 N} \right) .  
\ee
We define $G(w)$ by $G(w)= w^{1-1/N} - 2(1- {1/N}) w^{1/2} + 1 - {2/N}$. 
The derivative of $G(w)$ is given by 
\be 
  G^{'}(w) = (1-1/N) w^{-1/2} (w^{1/2-1/N}-1) 
\ee
It is clear that $ G^{'}(w) < 0$ for $0 < w< 1$. 
Since $G(1)=0$,  we have $G(w) > 0$ for $0 < w< 1$.  Therefore 
we have  $F^{'}(w)  <0$ for $0 < w< 1$. 
Since $F(1)=0$, we have  
$F(w) > 0$  for $0 < w< 1$. Hence, we obtain inequality  (\ref{eq:ineq2}).  

%%%%%%%%%%%%%%%%%%%%%%%%%%%%%%%%%%%%%%%%%%%%%%%%
\setcounter{equation}{0} 
\renewcommand{\theequation}{C.\arabic{equation}}
\section{Proof of inequality  (\ref{eq:ineq-dx^2})}

Multiplying both hands side of (\ref{eq:ineq-dx^2})
by $\sin^2(2 \varphi/N)  \cos^5 \varphi/ \sin \varphi$ we show that 
 (\ref{eq:ineq-dx^2}) is equivalent to 
\be 
\cos^2(\varphi - \frac {2\varphi} N) -  \cos \varphi \cos(\varphi - \frac {2\varphi} N ) 
\frac {\sin 2\varphi} {N \sin(2 \varphi/N)} > 
\sin(\frac {2 \varphi} N) \sin(2 \varphi - \frac {2\varphi} N) \, .    
\ee
Making use of $\sin \alpha \sin \beta= (\cos(\alpha-\beta) - \cos(\alpha+\beta))/2$ 
and $\cos^2 \alpha = (1 + \cos 2 \alpha)/2$  in the right-hand side and left-hand side, 
respectively, and multiplying both hand sides by $\sin(2 \varphi/N)/\cos^2 \varphi$ 
we reduce it to 
\be 
\sin( \frac {2 \varphi} N)  > {\frac 2 N} \sin \varphi \cos( \varphi - \frac {2 \varphi} N) 
\qquad 0 < \varphi < \pi/2 \, .  \label{eq:ineq-reduced}
\ee   
By applying (\ref{eq:ineq-alpha})  in lemma \ref{lem:ineq-alpha} with $\alpha=2/N$ and 
making use of the fact that $1> \cos \varphi - 2 \varphi/N) > 0$   
we show the reduced inequality (\ref{eq:ineq-reduced}).

%%%%%%%%%%%%%%%%%%%%%%%%%%%%%%%%%%%%%%%%%%%%%%%%
\setcounter{equation}{0} 
\renewcommand{\theequation}{D.\arabic{equation}}
\section{Proof of inequality  (\ref{eq:ineq-dx^2-m})}

Multiplying both hands side of (\ref{eq:ineq-dx^2})
by $\sin^2( (2 \varphi + m\pi)/N)  \cos^5 \varphi/ \sin \varphi $ we show that 
 (\ref{eq:ineq-dx^2-m}) is equivalent to 
\bea 
& & \cos^2(\varphi - \frac {2\varphi + m \pi } N)  
-  \cos \varphi \cos(\varphi - \frac {2\varphi + m \pi } N ) 
\frac {\sin 2\varphi} {N \sin ( (2 \varphi + m \pi) /N)} 
\non \\ 
& & > \sin\left( \frac {2 \varphi + m \pi } N \right) 
\sin \left( 2 \varphi - \frac {2\varphi + m \pi } N \right) 
\quad \mbox{for} 
\quad {\frac {m \pi} {2(N-1)}} < \varphi < {\frac {\pi} 2}  \, .      
\non \\
\eea
Making use of $\sin \alpha \sin \beta= (\cos(\alpha-\beta) - \cos(\alpha+\beta))/2$ 
and $\cos^2 \alpha = (1 + \cos 2 \alpha)/2$  in the right-hand side and left-hand side, 
respectively, and multiplying both hand sides by $\sin((2 \varphi + m \pi)/N)/\cos^2 \varphi$ 
we reduce it to 
\bea 
& & \sin \left( {\frac {2 \varphi + m \pi} N}  \right )  > {\frac 2 N} \sin \varphi 
\cos \left( \varphi - {\frac {2 \varphi + m \pi } N}  \right) 
\non \\ 
& & \qquad \mbox{for}  \quad {\frac {m \pi} {2(N-1)}} < \varphi < {\frac {\pi} 2}  \, .  \label{eq:ineq-reduced-m}
\eea   
Here we show the following inequality:  
\bea 
& & \sin \left( {\frac {2 \varphi + m \pi} N}  \right )  > {\frac 2 N} \sin \varphi 
\cos \left( \varphi - {\frac {2 \varphi + m \pi } N}  \right) 
\non \\ 
& & \qquad \mbox{for}  \quad 0 < \varphi < {\pi}  \, .  \label{eq:ineq-reduced-m}
\eea   
We define function $f(\varphi)$ by 
\be f(\varphi) = 
 \sin \left( {\frac {2 \varphi + m \pi} N}  \right ) - {\frac 2 N} \sin \varphi 
\cos \left( \varphi - {\frac {2 \varphi + m \pi } N}  \right) 
\ee
By taking the derivative we have 
\be 
\frac {df} {d\varphi} = {\frac 4 N}(1 - {\frac 1 N}) \sin \varphi \sin \left( \varphi- \frac {2 \varphi + m \pi} N \right) \, . 
\ee
The derivative vanishes at $\varphi = m \pi/(N -2)$, and hence function $f(\varphi)$
has the lowest value at $\varphi= m \pi/(N-2)$. By putting 
.$\varphi= m \pi/(N-2)$ we have 
\be 
f(\varphi_{min} )=  \left( 1 - \frac 2 N \right) \sin\left( \frac {m \pi} {N-2}  
\right) \, .      
\ee
The right-hand side is positive for $m=1, 2, \ldots, N-2$ and vanishes for 
$m=0$ and $N-1$.   
We therefore obtain the following: 
\be 
 \sin \left( {\frac {2 \varphi + m \pi} N}  \right )  > {\frac 2 N} \sin \varphi 
\cos \left( \varphi - {\frac {2 \varphi + m \pi } N}  \right) 
\, \,  \mbox{for}  \, \,  0 < \varphi < \pi .  \label{eq:ineq-reduced-min}
\ee   

\section*{Acknowledgements}
The authors would like to thank Profs. J.-S. Caux and A. N. Kirillov for useful comments.   
The present study is partially supported by Grant-in-Aid 
for Scientific Research No. 15K05204. Some part of the present research is performed 
when participating in the YITP workshop: New Frontiers in Non-equilibrium Physics 2015, 
July 21st - August 23rd, 2015, YITP, Kyoto University.

\end{document}